\documentclass[12pt]{JHEP3}

\usepackage{amsmath}
\usepackage{amssymb}
\usepackage{graphicx}

\newcommand{\be}{\begin{eqnarray}}
\newcommand{\ee}{\end{eqnarray}}

\newcommand{\bn}{\begin{enumerate}}
\newcommand{\en}{\end{enumerate}}

\def\IC{\mathbb{C}}

\def\IP{\mathbb{P}}
\def\IR{\mathbb{R}}
\def\IZ{\mathbb{Z}}

\def\CM{{\cal M}}
\def\CN{{\cal N}}
\def\CO{{\cal O}}

\def\l{\lambda}
\def\m{\mu}
\def\n{\nu}

\def\hA{\hat{V}}

\newcommand{\vac}{\ket 0}

\def\tr{{\rm tr}}
\def\det{{\rm det}}

\newcommand{\ket}[1]{\vert  #1\rangle}
\newcommand{\bra}[1]{\langle #1 \vert}
\newcommand{\vev}[1]{\left\langle{#1}\right\rangle}

\title{ The BPS spectrum of monopole operators in ABJM: towards a field theory description of the giant torus. }

\author{David Berenstein$^{1}$ and Jaemo Park$^{2,3,4}$

\\
\\
$^1$Department of Physics, UCSB, Santa Barbara, CA 93106 , USA
\\
$^2$Department of Physics, POSTECH,
Pohang 790-784, Korea
\\
$^3$Postech Center for Theoretical Physics (PCTP), Postech, Pohang
  790-784, Korea
\\$^4$Department of Physics, Stanford University, Stanford, CA
94305-4060, USA
\\
\\
E-mail:
\email{dberens@physics.ucsb.edu, jaemo@postech.ac.kr}
}

\abstract{We study the BPS spectrum of monopole operators in ABJM
theory. First we work out the complete spectrum of the chiral ring
by using a semiclassical analysis of the field theory compactified
on a two sphere. By properly taking into account the full
quantization condition of monopole charges, we show that the
moduli space of ABJM theory with Chern-Simons level $k, -k$ is a
particular $\IZ_k$ cover of the symmetric product of
$\IC^4/\IZ_k$. As a byproduct of our analysis we show that the
dibaryon operators are gauge invariant and dual to D4 brane
giants. We also work out in detail the spectrum of fluctuations
around half-BPS monopole configurations and we find candidate
states for a dual BPS configuration to the giant torus solution
found by Nishioka and Takayanagi in the supergravity limit. We
also discuss more general BPS states. }

\preprint{SU-ITP-09/14}

\begin{document}

\section{Introduction}

The study of supersymmetric states in superconformal field theories
serves as a very useful bridge to connect the physics of field
theories at weak coupling with field theories at strong coupling. In
the case of the AdS/CFT correspondence \cite{M, GKP, W}, the weakly
coupled field theories can usually be described by using perturbation theory, while the strongly coupled regime can be described as a dual string or M-theory compactification on geometries of large size of the form $AdS_{d+1}\times X$, where the original superconformal field theory lives in $d$ dimensions.

For example, the supergravity spectrum of $AdS_5\times S^5$ \cite{KRN} was reproduced exactly by considering the spectrum of half-BPS  representations (operators) of ${\cal N}=4 $ SYM \cite{W}. However, the spectrum does not stop just at the level of the perturbative supergravity spectrum of the compactification. Some extended D-brane configurations can also  be BPS and can be described by similar constructions of free field gauge invariant operators (states). These can be dibaryon operators \cite{Witbar} and
giant graviton configurations \cite{McST}. Giant gravitons can also grow into the AdS space \cite{SGol, HHI}. The characterization of the dual states to giant gravitons at weak coupling in ${\cal N}=4$ SYM was developed in \cite{BBNS, CJR}, and this was reformulated as a free fermion quantum hall droplet system \cite{Btoy}. The fermion droplet configurations were further shown to be realized directly by bubbling geometry configurations \cite{LLM}. Also, non-trivial baryon chiral ring spectra were found to be correlated with the classification of topologically non-trivial brane fluctuations \cite{BHK}, finding a general class of matched states.
These developments show a rich interplay between the evolution of results from field theory and supergravity. They also range from a simplified perturbative analysis all the way to topology changing processes in gravity.

Recently, a new AdS/CFT pair system whose coupling constant can be tuned between weak and strong coupling has arisen in the study of superconformal field theories in three dimensions: the ABJM field theory \cite{ABJM}. This system provides a new opportunity to test the connections between string theory/M-theory on AdS spaces and quantum field theory on the boundary.  Because the study of the theory is in a new dimension and the dual string theory is type IIA string theory or M-theory, it has received wide attention. Finding how this case is different than previous situations is very interesting, as well as a potential source of new intuition on the gauge/gravity correspondence.

A particularly important new phenomenon in the quantum field
theory is the existence of non-perturbative operators in the
chiral ring, described by operators that insert monopole charges
\cite{ABJM}. The structure of these objects have been described in
various levels of detail in
\cite{BT,HLLLPY,KKM,Imamura:2009ur,Kim:2009wb,JS}. A more recent
analysis in Chern-Simons Yang Mills was carried out in \cite{BKK}.
These operators are necessary to describe the full moduli space of
vacua of the field theory, and they also play a role in the AdS
geometry, as the magnetic charge in the field theory is dual to
the D0 brane charge. Thus, they provide the essential ingredient
that is required to understand the origin of the eleventh
dimension of M-theory. The fact that the field theory is based on
$U(N)\times U(N)$ has important consequences for the rules of
quantization of magnetic charge \cite{KKM}.

In order to deal with such nonperturbative operators we have to deal with the conformal field theory in full generality.
In this approach, we can specify the conformal field theory by
enumerating all of the local operators and their correlation
functions. In this set up, the BPS operators are again playing a crucial
role in understanding the underlying  conformal field theory and extrapolating from weak to strong coupling.
The BPS operators which are non-perturbative  are annihilated by  suitable supercharge
generators of the corresponding superconformal algebras.

Since monopole operators are playing an important role in our
discussion we are using an equivalent definition of local operator insertions in
field theory in $R^3$, by dual corresponding states in $S^2 \times R$
via radial quantization of  the CFT. Thus enumeration of BPS operators is
reduced to working out the complete set of BPS states in $S^2 \times R$.
We apply this procedure in a two-fold way. First we work out the classical chiral ring in
the ABJM theory. We find that the classical system is characterized by configurations on the classical moduli space of vacua of the field theory with some constraints due to to classical flux quantization. A semiclassical quantization of these configurations leads to a collection of allowed quantum  numbers in the chiral ring. From here we can work out the exact chiral ring and thus we find the complete set of holomorphic order parameters that encode the
moduli
space of ABJM theory. This lets us specify the complete moduli space
of the theory, making a slight improvement over previous
descriptions. We recover that the moduli space is  a symmetric
product $Sym^N(\IC^4/\IZ_k)$ for the $U(M)\times U(N)$ theory for
$M > N$ as described in \cite{ABJ}.

However, our computation implies that the usual claim of the
moduli space of ABJM with the Chern-Simons level $k,-k$ as the
symmetric product of $\IC^4/\IZ_k$ should be modified. We find
that  the moduli space should be given by a $\IZ_k$ cover of the
symmetric product space $Sym^N(\IC^4/\IZ_k)$ for the $U(N)\times
U(N)$ ABJM theory. In the case of $N=2$, our result is very
similar to the moduli space computed in \cite{Distler:2008mk}, but
they have quotients by $\IZ_{2k}$ instead of $\IZ_k$. One can see
that our present setup makes the structure of the chiral ring
transparent. These techniques can be applied to many other SCFT's
in three dimensions. Such study is beyond the scope of the present
paper. This small modification does not change the basic picture
of M-theory. It is similar to what happens in the baryonic branch
of the conifold theory, where the geometry of moduli space has one
more complex dimension than a symmetric product. This is a $1/N$
modification of the dimension of moduli space. For a single brane,
the moduli space is $\IC^4$, but this is not carried over to all
branes. The rest of the branes see only the conifold geometry. The
associated Goldstone mode was found in the supergravity solution
in \cite{GHK} showing that the effect is visible in the
gravitational theory. In our case the modification does not grow a
full  complex dimension, but it grows the volume of moduli space
by a discrete amount. We could say that we grow a discrete
dimension if we want to. The important thing is that this
phenomenon will only affect one brane out of $N$, and the rest of
the branes will see the quotient $\IC^4/\IZ_k$. One nice
by-product of  our analysis is the clarification of the dibaryonic
operators which are dual to $M5$ branes wrapping torsion cycles.
With the previous understanding of the moduli space of ABJM
theory, the corresponding dibaryon operators would not  be
gauge-invariant. With our current understanding of the moduli
space, the dibaryon opeartors are indeed gauge-invariant. Dibaryon
operators in $N=4$ Chern-Simons theory dual to M5 branes wrapping
integer cycles are considered in \cite{Imamura:2008ji}.

We also identify the BPS opeartors of ABJM theory dual to interesting
BPS brane configurations in the gravity side.
From the point of view of gravity, a new phenomenon is given by BPS
brane configurations
with non-trivial topologies on the AdS geometry, as shown by Nishioka and Takayanagi \cite{NT}. They constructed a giant torus configuration on the gravitational side of the duality that preserves some fraction of the supersymmetry. The corresponding states carry a large amount of angular momentum and D0 brane charge. This setup poses a challenge for the field theory dual description of these states. Indeed, giant gravitons growing into AdS are usually thought of as going into the Coulomb branch of the theory and a corresponding Higgsing procedure \cite{HHI, Btoy} (see also \cite{BHart}) with some pattern of symmetry breaking (lets say $U(N)\to U(N-1)\times U(1)$).
If we consider a global coordinate system on $AdS$, where we have a global time, a radial direction and a two sphere,  a giant torus does not cover the sphere, but only part of it. Therefore this  Higgs mechanism can not happen everywhere uniformly. Moreover, one has that at fixed angle position on the sphere
one can find either two points of the torus, or none. So whatever is happening has to involve more than one brane and can only be described by configurations that involve some non-abelian degrees of freedom.  Therefore it is interesting to ask how these objects arise from the dual field theory. In this paper we address some of these issues by considering some special configurations that were studied in \cite{NT}. They showed that if one considers a set of D0 branes that are polarized into a fuzzy sphere, then there are BPS strings suspended between the north pole and the south pole which carry a set amount of angular momentum depending on the D0 brane charge, and that the condensation of these strings opens a funnel between the north and south pole.
In this paper we will show how to account for the quantum numbers of these strings and how the ingredients of the field theory work together to give a BPS configuration. We provide a basic starting point to address the giant torus configurations in field theory.

The content of the paper is as follows. In section 2 we review the
superconformal algebra for 3-dimensional field theories setting up
the basic notations for the future discussion. In section 3, we
work out the the chiral ring of ABJM theory, thereby working out
the underlying moduli space.  In section 4 we review the M2 dual
giant graviton configurations following \cite{NT}. Our main
interest lies in finding out the dual field opertors of their
supergravity solution of Fuzzy sphere and giant torus
configurations. Fuzzy sphere in $AdS_4 \times CP^3$ is also
considered in \cite{HM}. In section 5, we work out the dual field
theory configuration of Fuzzy spherical giant. This follows also
from the work \cite{JS}. In section 6, we find BPS configuration
for giant torus when the fundamental string number is not large.
In section 7, we make some comments about other more general BPS
states. We expect huge degeneracies for such general BPS states,
which might lead to the proper counting of black hole microstates.
This counting is beyond the scope of the present paper.

\section{The superconformal algebra for 3D field theories}

This section is mostly a review, so that the paper is self-contained. The main purpose of the review is to describe various aspects of the superconformal field theories in three dimensions. In particular, we want to pay attention to the unitarity bounds on superconformal representations, so that the BPS operators can be better understood in the later sections.

The superconformal algebra generators can be classified by the
following diagram, according to the
dimension of the corresponding operators
\begin{equation}
\begin{matrix}
-1&& & K_\mu&\\
\frac {-1}2 &{\quad} &&S_{\alpha}^i&\\
0& &M_{\mu\nu} & \Delta& R_{ij}\\
\frac 12&& &Q_{\alpha}^j&\\
1 && &P_\mu&
\end{matrix}
\end{equation}
The column on the left indicates the dimension of the corresponding
generator. The generators $S, K$ are the special
(super)-conformal generators respectively. $\Delta$ is the generator
of dilatations and $M$ are the generators of
rotations. $R_{ij}$ is the $R$ charge. If we have $\tilde N$
supersymmetries, the $R$-charge is $SO(\tilde N)$ and
the supercharges transform in the fundamental representation.

The unitarity of the representations is defined by making all
operators of dimension zero hermitian, while those of
opposite dimension being the hermitian conjugates of each other. This
is, we have $K_\mu = P_\mu^\dagger$,
and $(Q_\alpha^i) = (S^{i\alpha})^\dagger$. The spinors are doublets of $SO(3)$, the group of rotations.

 The commutation relations of interest to us are the ones between $Q$ and $S$. These are of the following form:
\begin{equation}
\{ Q^i_\alpha,S^{j\beta}\}= a  \delta^{ij} \frac 12 M_{\mu\nu}
\sigma^{\mu\nu}\!^\beta_{\alpha}+ b \delta^{ij}
\Delta \delta^\beta_{\alpha}+ c R^{ij} \delta^\alpha_\beta
\end{equation}
 The first two terms are symmetric in  $i,j$. This forces them to be
 proportional to $\delta^{ij}$ in order to be
$SO(\tilde N)$ invariant. These are decomposed into the two $SO(3)$ representations $\frac 12\otimes \frac 12= 0\oplus 1$. The representation with spin zero is $\Delta$, while the one of spin one is $M$. The term antisymmetric in $ij$ is given by the $SO(\tilde N)$ generators, and it should be of spin zero. The coefficients $a,b,c$ are fixed by
the Jacobi identities and the normalization given by the usual supersymmetry algebra $\{ Q_\alpha,Q_\beta\} \sim 2 P_\mu \sigma^\mu_{\alpha\beta}$. This gives us $a=b=c=2$.

We classify states in a unitary representation according to the dimension (the eigenvalue with respect to $\Delta$).

Consider now the following positive operator (under the unitary relations given by $S\sim Q^\dagger$)
\begin{equation}
\{S^{i\alpha}, Q_{i\alpha}\} \sim \Delta \geq 0
\end{equation}
with index summation over $\alpha, i$. This shows immediately that $\Delta\geq 0$. This is a unitarity bound. This shows that $S$ lowers the dimension of states, but it can not do so forever.
The superconformal primary states  (superprimary) $\ket \alpha$ are characterized by $S\ket{\alpha}=0$. They are minimum weight states of a representation and serve to characterize them completely.

Consider now the case of a single index $\alpha$ (no summation)
\begin{equation}
\{S^{i\alpha}, Q_{i\alpha}\}\sim a \delta^{ij} M_{\mu\nu} \sigma^{\mu\nu}\!^\alpha_{\alpha}
+ b \Delta
\end{equation}
This gives us an inequality between spin and the dimension. The $\sigma^{\mu\nu}$ are essentially Pauli matrices, and it is standard convention that the two spinor indices are classified according to the their $L_z$ value. Written in this form, for both choices of $\alpha$, we end up with the following pair of inequalities
\begin{equation}
b \Delta \pm a L_z \geq 0
\end{equation}
so that $\Delta \geq L_z$. Now, if we choose an $SO(2)\subset SO(\tilde N)$ and write the corresponding $Q,S$ in complex notation, we can evaluate the following anticommutators
\begin{eqnarray}
\{ Q_\alpha, \bar S^{\alpha}\} =   \Delta \delta^\beta_\alpha+  R\pm  L_z\geq0\\
\{ \bar Q_\alpha,  S^{\alpha}\} = \Delta \delta^\beta_\alpha-  R\pm  L_z\geq 0
\end{eqnarray}
So that the dimension of operators is greater than their $R$-charge plus their spin. The inequalities can only be saturated by states of maximum spin $L_z$. States that saturate these inequalities will be labeled BPS states. In the Hilbert space if the inequalities are saturated then some supersymmetries $Q$ necessarily act by zero on the corresponding superprimary state. This is phrased by saying that the highest weight state leaves some supersymmetries unbroken.

 To have a non-zero R-charge, we need at least $\tilde N=2$ superconformal invariance in three dimensions. This is the same number of supersymmetries as the four dimensional reduction of supersymmetry to three dimensions.
These relations are a consequence of unitarity. They are also called the BPS bounds of the corresponding superconformal field theory.

The simplest conformal field theory is a free (chiral) superfield $\phi$. The dimension of $\phi$ is one half, and the dimension of it's superpartner is one. The chirality condition makes it so that the lowest component $\phi$ is annihilated by one of the $Q$ operators, and therefore it saturates the unitarity bound.The R-charge in this case is $SO(2)$ and the superfield is of spin zero. This gives it an $R$-charge value of $1/2$. If we increase the supersymmetries and keep the theory free, (let us say to $\tilde N=4$), then the scalars are a spinor representation of $SO(\tilde N)$.

For the special case of the ABJM theory, it is convenient at this stage to specify the R-charge quantum numbers of the fields. The theory has a $U(N)\times U(M)$ gauge group and an SO(6)-R-charge. With respect to a natural ${\cal N}=2$ supersymmetry in three dimensions, we include the $SO(2)$ quantum numbers of the fields also. It contains some scalar fields $\phi$ and it's fermion superpartners $\psi$.
\begin{table}[ht]
\begin{tabular}{|c|c|c|c|c|c|}
\hline & $U(N)$ & $U(M)$ & $SO(6)_R$ & $SO(2)_R$&Spin\\
\hline $Q_{\alpha}$ & 1&1& 6 & $\pm1 \oplus 4\times 0$ & $\frac 12$\\
$\phi$ & $N$ & $\bar M$& 4 & $2\times \frac 12 \oplus 2\times \frac{-1}2$&0 \\
$\psi$ & $N$ & $\bar M$ & $\bar 4$ & $2\times \frac {-1}2 \oplus 2\times \frac{1}2$&$\frac 12$\\
$V_\mu^1$ & Adj & 1 & 1&0& 1\\
$V_\mu^2$ & 1 & Adj & 1&0& 1\\
\hline
\end{tabular}
\end{table}
It is natural to split the scalars into complex fields $A,B^\dagger$, such that $A$ has eigenvalues $1/2$ with respect to the $SO(2)_R$ charges, and $B^\dagger$ has charge $-1/2$. These are the lowest components of chiral superfields on the dimensional reduction of the ${\cal N}=1$ superspace in four dimensions. Their fermion partners are $\psi_A$ and $\bar \psi_B$ which have $SO(2)$ R-charges of $\frac{-1}2$ and $\frac 12$ respectively. The $A$ superfield mutiplets are chiral with respect to the ${\cal N}=2$ supersymmetry we have chosen, while the $B^\dagger$ are antichiral.

\section{The moduli space of vacua and the chiral
ring.}\label{section3}

We will reanalyze the moduli space of vacua of the ABJM model, for the case of $U(N)_k\times U(M)_{-k}$ theory, where $M\geq N$. In general, the moduli space for the ABJM theory with $N=M$ has been described by a symmetric product of $N$ particles on $\IC^4/\IZ_k$. However, for $N=M$, we will find that the moduli space is actually bigger by a discrete amount: the true moduli space is a $\IZ_k$ cover of this symmetric product. For $M>N$, the symmetric product $Sym^N(\IC^4/\IZ_k)$ is the correct description of the moduli space.
 To understand this discrepancy, we need to analyze the chiral ring of the theory in a lot of detail. In particular, this problem can only be solved by a proper understanding of the allowed magnetic monopole operators of the theory. This is related to the problem of finding the
 correct quantization conditions on the magnetic fluxes of the operator insertions. Many papers in the literature have missed this subtlety.

We should remember that the moduli space of vacua is the collection of vacua of a supersymmetric theory that preserve all of the supersymmetries. We will be working in theories with four supercharges in three dimensions. This is the same amount of supersymmetry as that for ${\cal N}=1$ supersymmetry in four dimensions. Hence, we can use the ${\cal N}=1$ superspace to make the analysis. The theory we will study in detail is the ABJM model, which has more supersymmetries, but it is instructive to keep the superspace description in this fashion as it allows for general statements to be made.

The supercharges are given by standard superspace expressions
\begin{equation}
Q_\alpha= \frac{\partial}{\partial\theta^\alpha} - i \sigma^\mu_{\alpha\dot\beta} \bar\theta^{\dot\beta} \partial_\mu
\end{equation}
And the covariant derivatives are given by
\begin{equation}
D_\alpha= \frac{\partial}{\partial\theta^\alpha} + i \sigma^\mu_{\alpha\dot\beta} \bar\theta^{\dot\beta} \partial_\mu = Q_\alpha+ 2\sigma^\mu_{\alpha\dot\beta}\bar\theta P_\mu
\end{equation}
where we are identifying $P_\mu= i\partial_\mu$ as the generator of translations.

We want to think of $Q$ and $P$ as operators in the Hilbert space of states.
Assume that we have  a supersymmetric vacuum $\vac$. We thus have that
\begin{equation}
P \vac= Q\vac = \bar Q\vac=0
\end{equation}
The different vacua can be parametrized by order parameters that distinguish them. The order parameters are local gauge invariant operators in the field theory. Thus, vacua are parametrized by vacuum expectation values of superfields
\begin{equation}
\bra 0 \CO(x, \theta, \bar\theta) \vac
\end{equation}
Translation invariance of the vacuum implies that
\begin{equation}
\bra 0 [P_\mu,\CO(x, \theta, \bar\theta)] \vac = 0 =i\partial_\mu \bra 0 \CO(x, \theta, \bar\theta) \vac
\end{equation}
so that the vacuum expectation values are invariant under translations. Lorentz invariance implies that $\CO$ is a scalar field, so it is a bosonic superfield.

 Similarly, we
can use the supersymmetry of the vacuum to show that
\begin{equation}
\bra 0 D_{\alpha} \CO(x, \theta, \bar\theta) \ket 0
= \bra 0 [ Q_{\alpha}+ 2\sigma^\mu_{\dot\alpha\dot\beta}\bar\theta P_\mu,\CO(x, \theta, \bar\theta) ]\vac =0= D_{\alpha} \bra 0 \CO(x, \theta, \bar\theta) \vac
\end{equation}
In this equation $Q,P$ are operators, while the $\theta, \bar\theta$ are parameters  and therefore $Q,P$ do not act on them.
Similarly we can show that $\bar D \bra 0 \CO(x, \theta, \bar\theta) \vac=0$.

This shows that on the moduli space of vacua the superfields that can get vacuum expectation values are both chiral and antichiral (they are annihilated by both $D, \bar D$). If the operators are already chiral, the antichirality of the vacuum expectation value is a property of the vacuum solution, but not of the general correlator of operators.

Moreover, if a superfield can be written as
\begin{equation}
\CO(x, \theta, \bar\theta)= \{D_\alpha, G(0,\theta,\bar\theta)\} \label{eq:Dcoh}
\end{equation}
then the same type of manipulations as done show that the expectation value of $\CO$ vanishes.

This is why the order parameters of a vacuum manifold can be
parametrized by equivalence classes of chiral operators: the
cohomology of $\bar D$. Also, the product of two chiral operators at
different positions is chiral and independent of the position. This is,
\begin{eqnarray}
\bra 0 \CO(x_1, \theta, \bar\theta) i\partial_{x_2}\CO(x_2, \theta, \bar\theta) \vac&=&
\bra 0 \CO(x_1, \theta, \bar\theta) [P,\CO(x_2, \theta, \bar\theta)] \vac\\
&=&\bra 0 \CO(x_1, \theta, \bar\theta)[\{D_\alpha, {\bar
  D_{\dot\alpha}}\},
\CO(x_2, \theta, \bar\theta)] \vac  \nonumber \\
&=& \bar D_{\dot\alpha} \bra 0 \CO(x_1, \theta, \bar\theta)[D_\alpha, \CO(x_2, \theta, \bar\theta)] \vac
 \nonumber \\ &=&0.   \nonumber
\end{eqnarray}
The last line follows from a simple generalization of
($\ref{eq:Dcoh}$) for the case
where we have many $x$ coordinates. The upshot is that vacuum
expectation values of products of chiral
operators are independent of the positions of the insertions. The Cluster decomposition principle guarantees that the expectation values can be computed in the limit where we separate the operator insertions infinitely away from each other: the vacuum expectation values factorize. And thus the operators that belong to the cohomology of $\bar D$ have a ring structure in their vacuum expectation values. This is the chiral ring of the theory. Here we are paraphrasing the arguments found in \cite{CDSW} applied to a three dimensional setup.

Having the chiral ring gives the vacuum manifold a complex structure: the vacuum expectation values of the chiral ring operators are complex coordinates parametrizing the vacua. Since the algebra of vacuum expectation values of the chiral ring is commutative, we find that the moduli space is characterized exactly by (one-dimensional) representations of this chiral ring algebra. This is the main reason why supersymmetric vacua are well described by algebraic geometric structures.

If we consider a conformal field theory, the chiral ring operators have additional structure. This is because the operators carry R-charge, and there is also a unitarity bound on their dimension versus the value of their R-charge (we have described this in a previous section of this paper). To be in the chiral ring, the lowest component of the operator can be chosen to saturate this unitarity bound and therefore the chiral ring operators can be identified with BPS operators: those that are annihilated by some of the $Q$ operators. Remember that $Q$ is of the schematic form
\begin{equation}
\bar Q_{\dot\alpha} \sim \bar D_{\dot\alpha} + \theta \sigma P
\end{equation}
So, on  a chiral superfield $\phi(x, \theta, \bar \theta)\sim \phi(x)+\theta\psi(x)+\dots$ we have that
\begin{equation}
[Q, \phi(x, \theta, \bar\theta)] \sim \bar D \phi +\theta \sigma \partial \phi \simeq \theta\sigma \partial \phi
\end{equation}
Since the right hand side does not have a lowest theta component we find that
\begin{equation}
[Q,\phi(x)]= 0
\end{equation}
so that if $\phi$ is a superprimary, it is annihilated by some of the $Q$ operators and it is therefore a BPS representation of the superconformal algebra.

Also, the unitarity bound on the operators guarantees that their OPE is nonsingular. Let us show this. Notice that
a general OPE is of the form
\begin{equation}
\CO(x_1) \CO(x_2) = \sum_{\Delta} |x_1-x_2|^{\Delta-\Delta_1-\Delta_2} \CO_\Delta(x_1)
\end{equation}
but the unitarity bound forces $\Delta\geq R_1+R_2$, while $\Delta_1=R_1$ and $\Delta_2=R_2$, so the OPE coefficient has a positive power law.
Thus, in the case of conformal field theories, the chiral ring is a graded ring and there is no OPE singularity of chiral ring BPS operators.

This implies that the moduli space of vacua is a cone. Given any vacuum, parametrized by graded coordinates (the ones that have fixed R-charge) we can rescale the coordinates as given by their R-charge. This rescaling can be done continuously and gives a $\IC^*$ action on the moduli space. One can take all vevs to zero (since all the R-charges of chiral ring operators are positive).

Notice that nowhere in the discussion above did we have to specify a lagrangian, or to declare that the operators $\CO$ were polynomials in the elementary fields.
This structure is universal, even in the presence of non-perturbative operators. For the case of three dimensional conformal field theories, elements of the chiral ring can be given by monopole operators that can not be written as polynomials in the perturbative fields. However, the chiral ring structure still allows that these generalized objects can be described by polynomials in elementary generators of some algebra (in the sense of algebraic geometry). In some cases these two descriptions are confused giving rise to inconsistencies in the  description of the physical system. Here, part of our objective is to clarify this issues precisely.

The idea now is to apply this information to the special case of the ABJM model. The main question that needs to be addressed is what is the list of quantum numbers of the allowed chiral ring operators. This list will give us the coordinate ring of some manifold and we can find out what the vacuum manifold is by knowing the exact structure of the chiral ring operators.

The additional advantage of Conformal Feld Theories is that local operator insertions at the origin are equivalent (via the Operator-state correspondence) to the spectrum of states for the same conformal field theory on the cylinder. Thus, for a $d$-dimensional conformal field theory we have an equivalence between operators $\CO(0)$, and states $\ket \CO$ for the theory compactified on $S^{d-1} \times \IR$. The time coordinate (represented by $\IR$ in the product manifold) is radial time.  The isomorphisms
of representations of the conformal group are handled in natural way. This is,
commutators with the superconformal charges in the operator version, are taken to the natural
action of the supercharges on the Hilbert space of states.

As such, we have necessarily that descendants are mapped to descendants
\begin{equation}
[P_u, \CO(0)] \to P_\mu \ket\CO
\end{equation}
and that the generator of dilatations, which measures the scaling dimension of operators
\begin{equation}
[\Delta,\CO(0)]= \Delta_{\CO} \CO(0)
\end{equation}
 gets mapped to the time evolution under radial time, so that
 \begin{equation}
 \Delta \to H
 \end{equation}
where $H$ is the hamiltonian on the sphere. Similarly, the $R$ charge quantum numbers get mapped to the same exact $R$-charge quantum numbers on the operator side.

The Hermiticity conditions for the theory on the sphere are given by $Q_\alpha^\dagger= S^\alpha$, etc. These are the hermiticity conditions that we used in the previous section. The unitarity bounds are
\begin{equation}
\{Q, S\} =2( \Delta\pm R \pm L_z) \geq 0
\end{equation}
these relate the energy, the R-charge and the spin of the operator. Standard operators in the chiral ring have $L=0$ and $\Delta=R$. Thus, they are spherically symmetric and their energy is saturated by the R-charge.

Because the field theory on $S^2$ is on a compact space, the spectrum of states is discrete. Also, energy does not dissipate, so many states can be described by classical solutions to the equations of motion. In order to quantize these solutions, one can use semiclassical quantization techniques (in particular Bohr's rule) by requiring that various action angle variables have integer periods in units of $\hbar$. Here, we will adopt a scheme that is better adapted to the problem we are studying.

From our previous discussion, it follows that monopole operators in the chiral ring get mapped to states in the Hilbert space with non-trivial magnetic fields on the $S^2$. Their $R$-charges and energy can be measured by studying (semi-) classical
solutions of the field theory on the sphere that are compatible with the classification of fluxes of a $U(M)\times U(N)$ gauge connection.

Now, we want to repeat some of the arguments of \cite{Bcon} to
describe such classical solutions (this is a generalization of the
ideas of \cite{BlargeN} for the case of $\CN=4$ SYM in four
dimensions). Some similar ideas have also been presented in
\cite{Grant:2008sk} for more general states in ${\cal N}=4 $ SYM in four dimensions.
Such a program for the case of the ABJM  model was sketched in \cite{BT}, but some errors
were made in the description of the correct flux quantization conditions, so some elements of the chiral ring were missed. For this paper we need more details: we need the complete classical solutions with all the vacuum expectation values normalized accordingly. These solutions will be needed in future sections as background field with respect to which we will quantize in small fluctuations of other fields.

Our starting point is the action for the ABJM theory. The action has
various pieces. It is easiest to follow the work of Benna et al. to
describe the action \cite{BKKS} in terms of four dimensional $\CN=1$ superfields. The field content can be described in terms of two $\CN=2$ vector multiplets in four dimensions for the gauge groups $U(N)$ and $U(M)$, and two hypermultiplets $H_{1}$ in the $(N, \bar M)$, $H_2$ in the $(\bar N, M)$ representation of the gauge field. The action for the hypermutiplets is the standard dimensional reduction of the action for $\CN=2$ matter in four dimensions down to three. For the vector multiplets, however, the action is of Chern-Simons form. This means that all of the degrees of freedom of the vector mutiplet are auxiliary fields. An $\CN=2$ vector multiplet splits under $\CN=1$ superspace into a vector real superfield plus a chiral superfield $\phi$ in the adjoint. We have two such chiral superfields, one for $U(N)$ and one for $U(M)$ connnection.
Similarly, each hypermultiplet splits into a chiral and an antichiral superfield (which after complex cojungation becomes a second chiral superfield with the opposite gauge quantum numbers). Let us call the split of $H_1\to A_1, B_1$, and $H_2\to B_2,A_2$, where both of the $A,B$ are chiral.

The field $\phi$ couples to $A_1, B_1$ and $A_2,B_2$ via a superpotential, which gives a term in the lagrangian of the form
\begin{equation}
\int d^2 \theta\tr(\phi_1 (A_1 B_1-A_2B_2)-(B_1A_1-B_2A_2)\phi_2)
\end{equation}
while there is no kinetic term for $\phi_1,\phi_2$. Instead, they are auxiliary fields with a quadratic action that is proportional to a mass superpotential
\begin{equation}
\frac k 2\int d^2\theta\left( \tr(\phi_1^2) - \tr(\phi_2^2)\right)
\end{equation}
Integrating $\phi$ out, gives us that
\begin{equation}
k\phi_1 =  (A_1 B_1-A_2B_2)
\end{equation}
and that
\begin{equation}
-k\phi_2= (B_2A_2-B_1A_1)
\end{equation}
Replacing these values in the action, gives a superpotential term of the form
\begin{equation}
\int d^2\theta \frac 1k \left(\tr( A_1B_1A_2B_2)-\tr( A_1B_2A_2B_1)\right)
\end{equation}
which is the same superpotential as that one of the conifold of Klebanov and Witten \cite{KW}.

Similarly, the dimensional reduction of a vector field to three dimensions contains an additional real scalar $\sigma$ in the adjoint, apart from the connection in three dimension. This is from the fourth component of the gauge field, which becomes a scalar. Its contribution to the potential for the scalars is of the form of a square
\begin{equation}
|[\sigma,A_1]|^2
\end{equation}
which arises from the dimensional reduction of the field theory kinetic term from four dimensions to three. The commutator expression is given by
\begin{equation}
[A_1,\sigma] = \sigma_1 A_1- A_1 \sigma_2
\end{equation}
so that the group contractions make sense.

The kinetic action of $\sigma$ is auxiliary, and it arises as a term of the form
\begin{equation}
k \int d^4 x \tr(\sigma D)
\end{equation}
where $D$ is the auxiliary field of the vector multiplet. The equation of motion of $D$ makes
$\sigma$ into a composite field:
\begin{equation}
k \sigma_1= A_1A_1^*+A_2A_2^*-B_1^* B_1-B_2^*B_2
\end{equation}
and similarly for $\sigma_2$. It is simplest to keep the $\sigma$ in the discussion.

In addition to these terms, the action contains a Chern-Simons term for the gauge fields, with levels $k,-k$ respectively.
Finally, the scalars are conformally coupled to the  background metric, so when we write the
theory on an $S^2\times \IR$ geometry, we get an extra contribution to the mass of the scalar fields from the background curvature. This mass squared gets normalized to $1/4$ on a unit sphere.

We now want to solve the equations of motion for states that saturate the BPS equality $H=R$.
The Hamiltonian $H$ is the Legendre transform of the Lagrangian, with all auxiliary fields integrated out (including $\sigma, D$ and the F-terms for the $A,B$ superfields). Also, the time component of the connection is used as a Lagrange multiplier that imposes the Gauss law constraints. We can furthermore choose a gauge connection where $D_t=\partial_t$.

The Legendre transform of the Chern-Simons action vanishes. This is because it is a first order action. It can also be understood from the fact that the Chern-Simons system on its own is a topological quantum field theory, so the Hamiltonian vanishes because the Chern-Simons  fields do not couple to a background metric.

The potential for the scalars is of the form
 of a sum of squares:
\begin{equation}
\int_{S^2} d^2\sigma \frac 14 |A|^2+\frac 14 |B^2| + |W_A|^2+|W_B|^2+
|[A,\sigma]|^2+|[B,\sigma]|^2  \label{scalarpotential}
\end{equation}
where the integral is over a single time slice.

Finally, we also have that the kinetic term is of the form
\begin{equation}
\int_{S^2} d^2\sigma |\pi_A|^2+|\pi_B|^2+ |\nabla A|^2+|\nabla B|^2
\end{equation}
where the $\nabla$ indicate covariant derivatives on the sphere with respect to the gauge connection and the metric. The gauge condition $D_t= \partial_t$ appears in the equation of motion of $A$ in Hamiltonian form
\begin{equation}
\dot A_1 = \pi_{\bar A_1} = \{ H,A_1\}_{PB}
\end{equation}
where we have the canonical Poisson brackets obtained from the gauge fixed degrees of freedom.

We have picked the $R$-charge which is compatible with this $\CN=1$ superspace, which is a subgroup of $SO(6)$, the R-charge group of the ABJM theory.

The corresponding $SO(2)$ R-charge is given by the following generator on the classical system
\begin{equation}
Q_R= i\int_{S^2}\frac12 [\pi_A A-\bar A\pi_{\bar A}+ \pi_B
B-\pi_{\bar B} \bar B]   \label{Rcharge}
\end{equation}
Remember that the scalars are in a $4$ dimensional spinor representation of the $SO(6)$ R-charge of the full ABJM theory. Hence, the $R$-charge of the fields must be normalized to one half, as is appropriate for spinors. This is also the canonical dimension of scalar fields in three dimensions.

Classically, only bosons get vacuum  expectation values, so we can ignore the fermions for this discussion. Also, we have the BPS inequality for all quantum states that shows that
\begin{equation}
H\geq Q_R
\end{equation}
and this is valid for all classical states as well (after all they are just coherent states of the quantum system). Moreover, we have that their Possion brackets vanish
$\{Q_R,H\}_{PB}=0$, which just states that the $R$ charge is a constant of motion.

Finally, if we consider the Hamiltonian function $\tilde H= H-Q_R$, the set of configurations that are a global minimum of $\tilde H$ is exactly the set of configurations that satisfy $H-Q_R=0$.
For such configurations we must necessarily have that
\begin{equation}
\delta_{A,\pi} \tilde H =0
\end{equation}
and that the Poisson brackets of all variables with $\tilde H$ vanish for these configurations (this is the standard statement that a global minimum of a Hamiltonian is stable and provides a time independent solution of the equations of motion). Thus, for any such configuration we have
that
\begin{equation}
\{F,\tilde H\}_{PB}|_{\tilde H=0} =0 =\{F, H\}_{PB}|_{\tilde H=0} -\{F, Q_R\}_{PB}|_{\tilde H=0}
\end{equation}
so that the usual equations of motion can be given by
\begin{equation}
\dot F = \{ H, F\}_{PB} = \{Q_R,F\}_{PB}
\end{equation}
These can be applied to the fundamental fields, giving
\begin{equation}
\pi_{\bar A}=\dot A= i \frac 12 A; \pi_{\bar B}=\dot B= i\frac 12 B
\end{equation}
and the gauge field is time independent (as it does not carry any $R$-charge). These are the simplified equations of motion for BPS configurations (naturally they are first order).

From here, it is easy to evaluate the specific value of $Q$ for these configurations,
\begin{equation}
Q= \frac 12 \int_{S^2} \left( |A|^2+|B|^2\right)
\end{equation}
Similarly, we can compute the value of $H$ using the simplified equations of motion, which is
\begin{eqnarray}
H &=& \int_{S^2} 2( \frac 14 |A|^2+\frac 14|B|^2)  \nonumber\\
&&+\int_{S^2} ( |\nabla A|^2+|\nabla B|^2) \nonumber\\
&&+\int_{S^2}  |W_A|^2+|W_B|^2+ |[A,\sigma]|^2+|[B,\sigma]|^2
\label{energy}
\end{eqnarray}
The expression $H-Q$ is seen to be given by a sum of squares (all the terms from the second line  onwards). All of these have to vanish to have an equality of $Q_R,H$.

This shows that the solutions must satisfy $\nabla A= \nabla B=0$, this is, $A,B$ are covariantly constant on the sphere. From the Chern-Simons equations of motion, this implies that the gauge field is also covariantly constant, so that the solutions are necessarily spherically symmetric. Moreover, we find that these constant field solutions must satisfy the conditions to be at a minimum of the potential of the field theory in flat space (all the squared terms have to vanish). These minimal energy conditions are exactly the ones that determine the classical moduli space of vacua.
Finally, notice that the velocities of the fields are determined from the fields themselves,
as the equations of motion are first order. Thus, only the complex values of $A,B$ need to be determined,
and the conjugate momenta follow.

We find a connection between classical BPS states on the sphere and the moduli space of vacua.
This should be a general feature for superconformal field theories with this amount of supersymmetry.
In particular, any classical BPS state on the sphere can be put into one to one correspondence with points
in the moduli space of vacua. After all, the global gauge symmetries can be used to show that only the
equivalence class of solutions up to global gauge transformations constitutes a solution.

We are still not done. We have solved the equations of $A,B$, but we also need to solve  the equations of motion for the gauge connection. We have simplified the analysis considerably  because we know that the solution is necessarily spherically symmetric, and that the gauge connection (in the gauge we have chosen) is time independent. Here, we can follow precisely the classical analysis done in \cite{BT}, so we will not repeat it in full detail.

The upshot is that the gauge field is covariantly constant for $U(M)\times U(N)$. This can be parametrized by a diagonal magnetic flux on the sphere for the $U(M)\times U(N)$. Since the scalar fields $A,B$ are spherically symmetric, this can only happen when the flux in the corresponding eigenvalue of $U(N)$ is matched to the magnetic flux on the eigenvalue of $U(M)$, and the $A,B$ are scalar diagonal matrices connecting between these eigenspaces of the same flux. Finally, the magnetic flux on a given eigenvalue can be determined by the gauge field equation of motion
\begin{equation}
k\phi_{1ii} = \int ({|A_{ii}|^2-|B_{ii}|^2}) = k\phi_{2ii}
\end{equation}
where the $\phi$ are the integrated flux over the sphere in an appropriate normalization. We will describe the normalization of these fluxes further on.

The whole system reduces to diagonal matrices, and the problem can be solved in $1\times 1$ blocks at a time, which means that we get $N$ copies of the $U(1)\times U(1)$ theory moduli space. Also, since the permutation of eigenvalues is a gauge symmetry, the moduli space looks like a symmetric product over $\IC^4$. However, we have to be more careful. The reason is that the flux quantization conditions on these solutions modify the analysis.

The key observation that we have now where we differ from the analysis \cite{BT} is on what the quantization conditions for the $\phi$ are. Consistency of the matter coupled to the double gauge connection between two eigenvalues requires a Dirac quantization on the difference of fluxes. Thus
\begin{equation}
\phi_{1 ii}-\phi_{2jj} \in \IZ
\end{equation}
for all $i,j$. If $M\neq N$, then some of the fluxes on $\phi_2$ vanish, and we get that the flux on each eigenvalue of $U(N)$ and $U(M)$ is quantized (as expected from the classification of such connections by Atiyah and Bott \cite{AB}). However, if $N=M$, then we find that the fluxes need not be integer, and so long as their differences are integer values, there is a consistent description of the dynamics.

You should also notice that for these spherically symmetric solutions $A,B$ are constant on the sphere. We will parametrize the vacua by variables
\begin{equation}
A^{1,2} \simeq \begin{pmatrix} a_1^{1,2}&&\\
&a_2^{1,2}&\\
&&\ddots
\end{pmatrix}, B^{1,2}= \begin{pmatrix} b_1^{1,2}&&\\
&b_2^{1,2}&\\
&&\ddots
\end{pmatrix}
\end{equation}
and similarly for $\pi_A, \pi_B, \pi_{\bar A}, \pi_{\bar B}$. We choose the $a,b$ to have a normalized kinetic term of the form $|\dot a|^2$, etc. This fixes the normalization relations by the volume of the sphere.
We furthermore have the
equalities given by $\pi_{\bar a_i} = \frac i2  a_i$, etc. These follow from the BPS constraints
on the equations of motion.

The symplectic structure $\omega=d \pi_A\wedge dA+ d\pi_B\wedge dB+\dots$ has a non-trivial pullback to the set parametrized by the $a_i, \bar a_i$. A straightforward computation
shows that it is proportional to
\begin{equation}
i d\bar a_i \wedge d a_i+i d\bar b_i \wedge d b_i
\end{equation}
Thus, on the space of BPS initial conditions (parametrized by the variables $a_i$ and $b_i$ and their complex conjugates), we have that $a$ and $\bar a$, and $b, \bar b$ are canonically conjugate to each other, and their symplectic structure is the Kahler form of $(\IC^4)^N$.
Quantization is straightforward. We need to choose a polarization. Given that the $a_i,b_i$
commute with each other under Poisson brackets, we find that all the wave functions can be written as wave functions that depend only on $a_i, b_i$. This will give us a holomorphic quantization of the moduli space.  Their canonical conjugate variables can be represented by
\begin{equation}
\pi_a \sim \partial_a
\end{equation}
For  the wave function to be single valued, it must be a power series of the $a,b$ variables.
The problem is finally solved by requiring some inner product that is compatible with the symmetries:
\begin{equation}
\bra {g(a,b)} \ket {f(a,b)} = \int da d\bar a db d \bar b \mu(a,b,\bar a, \bar b)
g^*(\bar a, \bar b) f(a,b)
\end{equation}
where $\mu$ is at this stage an unknown measure factor. We require it to be invariant under
the $SO(2)$ R-charge and it should be homogeneous. This measure can be calculated or guessed for many examples (see \cite{BlargeN,BHart,BT}).
This issue is beyond the scope of the present paper. This quantization of the BPS states gives us the holomorphic quantization of the classical  moduli space of vacua of the field theory.

What we need now are the consistency conditions from the flux quantization of this description.
We find that in these variables
\begin{equation}
Q_R = \frac 12(a \partial_a + b\partial_b)
\end{equation}
as is appropriate to have $a,b$ with R-charge  $1/2$. $Q_R$ essentially measures the degree of the polynomial on $a,b$. In a holomorphic quantization of a harmonic oscillator, this is the level occupation number of the oscillator.

We also find that
\begin{equation}
(k \phi_1)_{ii} = a_i\partial_{a_i} -b_{i}\partial_{b_i}=
k(\phi_2)_{ii} \label{cons1}
\end{equation}
and we know that $(\phi_1)_{ii}-(\phi_1)_{jj}$ is an integer.

For a general monomial in the $a_i, b_i$, $(a^1_i)^{n_{i1}}(a^2_i)^{n_{i2}}(b^1_i)^{m_{i1}}(b^2_i)^{m_{i2}}$, we get that
\begin{equation}
(k\phi_1)_{ii} = n_{i1}+n_{i2}-m_{i1}-m_{i2} \label{cons2}
\end{equation}
So that $\phi$ can be fractional in units of $1/k$. If one of the $\phi$ vanishes, then the $\phi$ is integer, and we get that the allowed values of the allowed monomials are given by
a constraint $ n_{i1}+n_{i2}-m_{i1}-m_{i2}=0\mod(k)$.

The variables $a^{1,2}_i,b^{1,2}_i$ represent a $\IC^4$. If we consider the $\IZ_k$ action
by $a_i\to \xi a_i$, $b_i\to \xi^{-1} b_i$, where $\xi^k=1$, we find that the allowed polynomials in $a,b$ described above are exactly those that are $\IZ_k$ invariant. Thus, we would find that the wavefunctions are single-valued on $\IC^4/\IZ_k$. Hence, the chiral ring wave functions can only differentiate many particles on $(\IC^4/\IZ_k)^N$, but not on $\IC^4$.

Also, the fact that
the eigenvalues being permuted is a residual gauge symmetry of the original system  implies that the allowed wave functions have to be symmetrized on the $i$ labels. Thus, the natural candidate for the wave functions, is that they are counted by the polynomial ring of invariants of
$\CM= Sym^N(\IC^4/\IZ_k)$. This object is the moduli space that has
been described in the literature
so far \cite{ABJM}. This is correct for the case of $M > N$ where one of the fluxes on $U(M)$ necessarily vanishes. These functions can all be generated by traces and their products.

However, for $N=M$ the analysis is more subtle. Remember that $\phi$ can be quantized in
fractional units, but if one such flux is fractional, all of them are. We say that
\begin{equation}
\phi_i \equiv \phi_j \mod(1)
\end{equation}
We find this way that
\begin{equation}
 n_{i1}+n_{i2}-m_{i1}-m_{i2}- (n_{j1}+n_{j2}-m_{j1}-m_{j2}) =0 \mod(k)
\end{equation}
This is the same ring of invariants of the space
\begin{equation}
(\IC^4)^N/(\IZ_k^{N-1})
\end{equation}
where the $\IZ_k^{N-1}$ are generated by comparing to the first eigenvalue,  $a_1\to \xi a_1, b_1\to \xi^{-1} b_1$, and $a_j \to \xi^{-1} a_j, b_j \to \xi b_j$. This is a $\IZ_k$ cover of $[(\IC^4)/(\IZ_k)]^{N}$. The permutation of eigenvalues still acts on this space, so the correct moduli space is given by
\begin{equation}
\CM = \left[(\IC^4)^N/(\IZ_k^{N-1})\right]/S_N
\end{equation}
Notice that this is not a symmetric product. It is a $\IZ_k$ cover
of a symmetric product. From here, the set of chiral ring states
is bigger than that for a symmetric product. The description of
the space as a quotient gives an explicit description of the
chiral ring, although counting states is rather involved. For
$N=2$ we get a similar result as the one found in
\cite{Distler:2008mk} for the $SU(2)_k\times SU(2)_{-k}$ theory.
The discrete group quotient is different. We get the $D_k$
quotient, whereas their result has a $D_{2k}$ quotient. Finally
for $M=N=1$ the moduli space is simply given by $\IC^4$ (it is an
obvious $\IZ_k$ cover of $\IC^4/\IZ_k$). For this case,  gauge
invariant monople operators are worked out explicitly in
\cite{HLLLPY}. The crucial point is that the gauge invariant
operator is the product of the flux creation opeartor and the
charge creation operator, which captures nicely the Gauss Law
constraint eq. (\ref{cons1}) and eq. (\ref{cons2}). One can also
directly work out the moduli space as  in \cite{ABJM}. The usual
claim that the moduli space being $\IC^4/\IZ_k$ depends on the
integer quantization condition  of the flux associated with
$F=dA_1+dA_2$ where $A_1, A_2$ are the gauge fields of two U(1)s.
The important point is that no matter is charged under $A_1+A_2$
and the integer quantization condition does not have to be
imposed. The only constraint comes from the Gauss Law constraint
and the flux of $F$ can be fractional in unit of $\frac{1}{k}$.
This changes the periodicity of the conjugate variable of
$A_1+A_2$ by factor $k$ compared with \cite{ABJM}, which leads to
the moduli space $\IC^4$.

The simplest polynomial we can find that is not in the symmetric product of the quotient is
\begin{equation}
a_1^1 a_2^1 \dots a_N^1 \sim \det (a^1)
\end{equation}
This state has the familiar quantum numbers of a dibaryon. It has magnetic constant flux on all of the eigenvalues, but no field degree of freedom sees the magnetic field. For the special case of $N=2$, $k=1,2$, these objects have been discussed by Klebanov et al \cite{KKM}.

We also see that $\det((a^1)^k)$ does belong to the symmetric product space, since $(a^1)^k$ is an allowed monomial for a wave-function on $\IC^4/\IZ_k$. This is how products of $k$ operators
with $\phi=1/k\mod(1)$ end up in the symmetric product.
This subtlety with the flux quantization was also missed recently in \cite{Kim:2009wb,JS}, where a more restricted semiclassical discussion of  BPS states was done. This quantization condition was studied correctly in \cite{Park:2008bk, Imamura:2008ji}, but the analysis here is exhaustive.

Since the moduli space is not a symmetric product, this means that the counting of chiral ring states by the Plethystic exponential is incomplete. One misses all states that have
$\phi\neq 0 \mod(1)$. We will call the value of $k\phi \mod(k)$ the $k$-ality of a state.
The quantization of the symmetric product space corresponds to the ring of operators with vanishing k-ality.

Notice also that the natural ring structure on the chiral ring is given at the level of wavefunctions by multiplications of polynomials. This is how one gets the standard relations that are required for adding charges of operators under OPE expansions. This is why we can identify the set of polynomials we found here with the chiral ring itself.

However, we want to reiterate that the presence of the magnetic field flux guarantees that
the corresponding states are not polynomials in raising operators of the fundamental fields on the sphere and should not be assigned operators of the form $\tr (A^k)$. This is an abuse of notation that captures the $R$-charge of the operator but makes no sense from the point of view of gauge invariance and leads to potential confusion. Even supplementing expressions of this type with 'monopole insertions' and 'Wilson lines' is not useful in the non-abelian setup, since one can not take a trace in any meaningful way. The use of such notations should be avoided.

 The description we have found should be interpreted instead as complete solutions to the non-linear
 equations of motion and have quantization conditions already at the classical level, because the gauge
 field fluxes are discrete at the classical level already: this is a classical result for covariantly
 constant field strengths on general Riemann surfaces \cite{AB}.

A note should be added here about possible choices in topology of
gauge field configurations. For example, in Yang Mills theory we
can ask that the global structure of the gauge group include the
center or not. This distinction does not appear at the level of
the lagrangian, but is specified separately and this changes the
spectrum  of magnetic monopoles of the field theory. Similarly in
this case of the ABJM model, we could require that only monopoles
with integer flux are allowed. This is the standard computation of
the chiral ring done in \cite{ABJM} that leads exactly to a
symmetric product. Both of these choices seem to be allowed at the
level of gauge theory. However, as was calculated in \cite{BT},
this choice of integer fluxes would forbid the dibaryon-like
operators above. The question to ask now is if this is visible in
the supergravity theory or not. The operators that distinguish
these choices are of large dimension (the energy is $N/2$), so
they can not be described by low energy supergravity computations
of the chiral ring. Instead, these operators have the natural
tension (conformal weight) associated to a D-brane in the
supergravity geometry. The dual objects are D4-branes wrapped on a
$\IP^2\subset \IP^3$ base of the type IIA supergarvity dual. These
are torsion cycles in the $S^7/\IZ_k$ quotient. A brane wrapping
these cycles would be forbidden if there is some type of discrete
torsion flux threading it. However, the natural setup one would
consider in gravity is that these extended objects are allowed
configurations, so the field theory dual must have operators with
the right quantum numbers. Since the presence of these operators
makes the chiral ring bigger for $N=M$, one must conclude that the
gravity dual knows about this. For $N\neq M$, these objects are
forbidden because $N-M$ is interpreted as RR-flux 4-form threading
the $\IP^2$. Thus the D4-branes would be anomalous: a number of
strings end on them and the corresponding operators are not gauge
invariant.

The situation we are presenting is very similar to the description
of field theories dual to cascading setups in four dimensions. In
those theories, for gauge groups  that are products of $SU(N)$,
the moduli space has a higher dimension  than the symmetric
product. The extra operators that count these directions are
dibaryons and one can talk about the baryonic branch of moduli
space. In these theories the baryonic symmetry (symmetries) is
broken for these configruations and one has  goldstone bosons in
the low energy effective theory. This was found directly in the
supergravity setup in one example in \cite{GHK}, for the baryonic
branch of the conifold. So one can argue that the supegravity
knows that the moduli space is bigger than a symmetric product.
This enhancement is seen when we consider the moduli space of a
single brane in the conifold. We get a $\IC^4$ moduli space rather
than a copy of the conifold. However, subsequent additions of
branes do not add one complex dimension for each brane, but only
one complex dimension total and the error one makes in considering
the moduli space as given by a symmetric product is of order $1/N$
and we find no trouble matching this result to considerations in
the dual gravity setup. In the ABJM setup, we only grow the moduli
space of the first brane by a discrete amount and all subsequent
branes get a $\IC^4/\IZ_k$ geometry. Thus the dual gravity
geometry is still the $AdS_4\times S^7/\IZ_k$ geometry, and is not
determined just by the first brane at the singularity.

The $Z_k$-fiber structure of the moduli space of the ABJM is
reminiscent of the discrete torsion of the orbifolds in the string
theory.  Due to the $B$-field in shrinking two-cycles, the moduli
space of D3-brane with the discrete torsion is entirely different
from that without the discrete torsion, even though the geometry
of the orbifold looks the same. In  \cite{Berenstorsion}, they
considered the geometry different from the standard $S^5/\Gamma$
space in that the singularities have monodromy of the resolving
spheres. The monodromy of the sphere makes the periods of the
twisted fields along the circle different from that of the
geometric circle, which changes the quantization of the masses of
the state. It is shown that this is nicely reproduced in the
D3-brane field theory.

Similar thing happens  for the monopole operators in ABJM case.
These are the operators detecting the M-theory circle. The problem
of which monopole charges are allowed should exist in the brane
realization of the theory. If the baryonic operators are allowed,
then even for a single brane, the expectation values of these
observables have monodromy on the orbifold geometry when going
around the M-theory circle. In field theory language, this is
reflected in the flux quantization condition of diagonal gauge
group,i.e., allowed fluxes can  be fractional in unit of $1/k$.
And this changes the periodicity of the conjugate variable of the
diagonal gauge field by the factor of $k$. That's how the above
phenomena of the monodromy is realized.

\section{Review of M2 giants}
In \cite{NT}  BPS states of M2 dual giant gravitons are constructed and
we review their constructions following their notations closely (additional results can be found in \cite{HMPS}) .
Schematically the metric for $AdS_4 \times S^7$ is given by
\begin{equation}
ds^2=\frac{R^2}{4} (ds_{AdS_4}^2+4d\Omega_7^2)
\end{equation}
where the $AdS_4$ metric is
\begin{equation}
ds_{AdS_4}^2=-(1+r)^2 dt^2 +\frac{dr^2}{1+r^2}+r^2 (d\theta^2 +sin\theta^2
d\phi^2)
\end{equation}
The $S^7$ can be written as
\begin{equation}
|z_1|^2+|z_2|^2+|z_3|^2+|z_4|^2=1
\end{equation}
and we parametrize $z_i$ as
\begin{equation}
z_i=\mu_i exp(i\xi_i).
\end{equation}
Let the conjugate momentum associated to $\xi_i$ direction be
denoted by $J_i$.
For $AdS_4 \times S^7/Z_k$ with the orbifold action $z_i \sim
e^{\frac{i 2\pi}{k}}z_i$ we have the additional identification
$\xi_i\sim \xi_i+\frac{2\pi}{k}$.
Upon dimensional reduction we have
\begin{equation}
ds_{S^7}^2=ds_{CP^3}^2+(dy+A)^2
\end{equation}
where $y$ is the diagonal direction of $\xi_i$ so that the angular
momentum $J$ along $y$ is given by $J_y=J_1+J_2+J_3+J_4$.
We are interested in two types of solutions.

\subsubsection*{ Fuzzy sphere in $AdS^4$}

Let us denote the worldvolume coordinates of a spherical M2(D2) brane as
\begin{equation}
\sigma_0\equiv\tau=t, \,\,\, \sigma_1\equiv \theta \,\,\,
\sigma_2\equiv \phi.
\end{equation}
An M2 'dual giant' configuration expanding spherically is given by
$y=\frac{t}{2}, r=r_0$
with the energy $E=\frac{P_y}{2}=\frac{1}{2}(J_1+J_2+J_3+J_4)$.
Notice that this configuration is moving in the $y$ direction. Upon dimensional reduction to $\IC\IP^3$,
the brane becomes motionless, but the momentum along the $M$-theory circle gets replaced by magnetic flux on the D2 brane.

If we take the $U(1)$ field strength $F=\frac{M}{2} \sin\theta d\theta
\wedge d\phi$ we obtain an induced D0 brane charge, so that one can check that the full energy of the D2 brane can be accounted for by a bound state of $M$ D0 branes with energy
\begin{equation}
E=\frac{kM}{2} \label{ejrelation}
\end{equation}
In ABJM side, this should be dual to operators whose conformal
dimension
is $E=\frac{kM}{2}$ and whose baryon charge is $kM$. Such a bound state in string theory would
usually look like a fuzzy sphere via the Myers effect \cite{Myers}.

\subsubsection*{Spinning dual giant gravitons}

In \cite{NT}  they further constructed another solution by introducing a nonvanishing
spin in $AdS_4$ which generically gives rise to  $\frac{1}{16}$ BPS states.
It has the functional form of $r=r(\theta)$ and $y=w\phi+\omega t$,
which means that it rotates in both $y$ and $\phi$ direction.
Note that $y$ denotes the 11-th circle direction and $\phi$ denotes
the angular direction in $AdS^4$.
Here $w \in Z/k$ is the winding number equivalent to fundamental
string charge while nonzero $\omega$ leads to the D0-brane charge after
the dimensional reduction. Remember that two-branes wrapped in the eleven dimensional circle get dimensionally reduced to fundamental strings.
One peculiar feature of the solution they found is that angular momentum along $\phi$ direction $P_{\phi}=S$ is related to
the angular momentum along $y$ direction $P_y=J=\Sigma_{i=1}^4 J_i$ via
$S=wJ$. Thus the total spin is proportional to the `number of strings'.

The energy is given by
\begin{equation}
E=S+\frac{J}{2}=\omega J=(w+\frac{1}{2}) J.  \label{etorus}
\end{equation}
The topology of the resulting membrane is shown to be a torus.
Upon the dimensional reduction, we have a bound state of D2-D0-F1
with $M$ D0-brane charge and $wk$ F-string charge with
\begin{equation}
J=P_y=kM,\,\,  S=P_{\phi}=wkM=wJ. \label{jsrelation}
\end{equation}
The resulting spinning D2 has nontrivial electric flux
$F_{t\theta}$ proportional to $w$
as well as magnetic flux $F_{\theta \phi}$ proportional to  $\omega$. In D2-brane systems the
electric flux corresponds to string winding number (it is the T-dual variable to momentum).

The torus can degenerate to a limit where a single string is suspended from the north pole to the south pole of a sphere. Such a  configuration can be thought of as a small perturbation of a fuzzy sphere by attaching one string to it. We will later be able to describe the duals of these states, though we have not found a simple way to describe the duals of the giant torus.

\subsubsection*{D4 brane giants}

One can also consider standard D4 branes wrapping the
$\IC\IP^2\subset \IC\IP^3$ and at the origin of AdS. These are
five-branes in the M-theory, and they wrap the torsion 5-cycle in
$S^7/\IZ_k$. Their dual realization corresponds to the dibaryon
operators. As we found in our description of the chiral ring,
these dibaryons carry a k-ality quantum number (the fractional
flux) that is valued in $\IZ_k$. This is the same $\IZ_k$ as the
torsion component of the homology cycles of $S^7/\IZ_k$ quotient.
It should be noted however, that if we choose to allow these
objects individually in gravity, we have to allow the
corresponding dibaryon operators in the field theory. These are
the duals of operators with fractional flux. Their absence would
indicate some kind of discrete torsion \cite{BT} that makes the
corresponding brane configurations anomalous. We already commented
on D4 brane and its dual operators in section \ref{section3}.

\section{ Fuzzy sphere Dual giant}

We can now describe the dual operators that must be matched to the Fuzzy Sphere giants. If we work at large values of $k$, as is appropriate for the type IIA string theory to be weakly coupled, then we can match to the semiclassical analysis we have done so far.

At large energies and R-charge we can replace the quantum results by  classical  values for the vev with specific time dependence (we can consider a coherent state if we want to). Thus we have that
\begin{equation}
A(t) = A(0) \exp(it/2)   \label{eqA}
\end{equation}
and similarly for $B$. This will give us a BPS state. The phase of $A(0)$ is specified by an initial condition, and the norm is calculated by equations (\ref{eq:vevs})
\begin{eqnarray}
 \lambda \vev{ |A_{1,2}|^2/4} =\frac 14 N_{A, 1,2} \nonumber\\
\lambda \vev{ |B|_{1,2}^2/4} = \frac 14 N_{B,1,2} \label{eq:vevs}
\end{eqnarray}
where $N_A$, $N_B$ denote the number of $a,b$ letters in our algebraic description (the A-charge and the B-charge). The constant $\lambda$ is a normalization constant: this is the volume of the sphere, if the $A,B$ have canonical normalization.  This is correct for a classical diagonal ansatz which makes the system look like a set of classically decoupled harmonic oscillators, with occupation numbers $N_A, N_B$.

For large vevs, we have states where the gauge symmetry is broken to
$U(1)^2 \times (U(N-1))^2$ (this is the commutant group of the
classical solution) below a scale characterized by the vevs. A large
vev on an eigenvalue is interpreted as a brane position even in global
AdS. The off-diagonal modes connecting the $U(1)^2$ and $U(N-1)^2$
become heavy due to the interaction potential in the full
theory. (Their mass grows as $v^2$ where $v$ is the vacuum expectation
value \cite{BT}.)
One can also argue that the radial dimension in moduli space corresponds to the radial direction of $AdS$ in global coordinates after a change of coordinates for very general setups \cite{BHart} by matching charges of BPS states with semiclassical positions from the saddle point approximation to a wave function.

A single D0 brane has baryon charge $k$ and dimension $k/2$. This must correspond to the simplest BPS monopole configuration that we have described in the ABJM theory, with one unit of flux. The map
of which operators get mapped to which D2 brane giant gravitons with flux was described in detail in \cite{JS}. The fuzzy sphere bound state we have described above in M-theory corresponds to all of the flux on the same eigenvalue. This can also be understood because a two brane of large size explores the radial direction. In the AdS geometry, the radial direction has a domain wall where the cosmological constant and the flux changes from having flux $N$ to having flux $N-1$, exactly at the location of the 2-brane. Such a shift is interpreted as the breaking of gauge symmetry $U(N)\to U(N-1)\times U(1)$ via a Higgs mechanism at the scale where the M2 brane is located \cite{HHI} (see also \cite{Btoy}). More general patterns of monopole fluxes can be mapped to various branes at various radial directions.
Notice the curious fact that the Fuzzy Sphere solutions' dual configurations are abelian, whereas the mechanism for making these objects in string theory requires a non-abelian configuration of D0 branes.

 We can also have a pure D2-brane with no D0 brane charge if we choose $N_A=N_B$. Such a brane moves in the reduction to type IIA theory. This is because the $\IC\IP^3$ coordinates of a collection of $A,B$ are given by
\begin{equation}
(w_1,w_2,w_3, w_4) \sim (A_1, A_2, B_1^\dagger, B_2^\dagger)
\end{equation}
so if  $A,B^\dagger$ have different time dependence and both are non-zero, the homogeneous coordinates $(w_1,\dots,w_4)$ will not be constant.

\section{Fluctuations on BPS Monopoles}

We have already described the dual operators to the the fuzzy sphere giant gravitons.
We now want to perturb these by adding `strings' to the D2-brane. Strings ending on a brane
are built in the field theory dual by suspending fundamental quarks from a `probe brane' and building the string and it's fluctuations from the non-abelian dynamics of gluons \cite{RYM}. In the operator language in CFT's, the spectrum of strings is usually built from a trace (spin chain) of fundamental fields. Here, we have already argued that the
fundamental field words are not good objects for describing monopoles. Hence, we should be very careful. Instead of this operator prescription, we have been using the description on the cylinder language, where we can understand how to deal with these issues by considering states in the full quantum field theory.  In this language, the spin chain is made of traces of raising operators with gauge indices contracted ( see \cite{Btoy,BCherkis} for a description of this dictionary). Now, we consider the ABJM field theory in the presence of a monopole background field (with arbitrary charge).

Since we can take the theory to be weakly coupled by making k large, we can do a perturbative expansion in the presence of background fields. The monopole operators are non-perturbative and therefore the vevs are large. They can induce large changes on the spectrum of quadratic fluctuations, but one can ignore backreaction and interactions between the fluctuations. If we want to add strings to a configuration of fuzzy spheres, we need to add quark excitations in the presence of the `probe' brane.
We will analyze the spectrum of quarks in the background field we are studying, ignoring the interactions between fluctuations.

Here we adopt the semiclassical approach and work out the quadratic fluctuation spectrum in
the presence of the monopole background.
We use the  Hermitian convention for the gauge fields
\begin{eqnarray}
D_{\mu}\bar{\phi}_A &=& \partial_{\mu} \bar{\phi}_A
+i(V_{\mu}\bar{\phi}_A-\bar{\phi}_A\hat{V}_{\mu}) \\
D_{\mu}\phi^A &=& \partial_{\mu} \phi^A+i(\hA_{\m}\phi^A-\phi_A V_{\mu})
\end{eqnarray}
Following the conventions of \cite{BT},  the  bosonic action on $S^2 \times R$
is given by
\begin{eqnarray}
S&=& \int_{S^2 \times R} d^3 x Tr (-D_{\mu} \bar{\phi}_A D^{\mu}
\phi^A -\frac{1}{4}\bar{\phi}_A\phi^A
-\frac{4\pi^2}{3k^2}(\phi^A\bar{\phi}_A\phi^B\bar{\phi}_B\phi^C\bar{\phi}_C
\nonumber \\
& &
+\bar{\phi}_A\phi^A\bar{\phi}_B\phi^B\bar{\phi}_C\phi^C+4\phi^A\bar{\phi}_B\phi^C\bar{\phi}_A
\phi^B\bar{\phi}_C-6\phi^A\bar{\phi}_B\phi^B\bar{\phi}_A\phi^C\bar{\phi}_C))
\nonumber \\
& & +\frac{k}{2\pi}\int_{S^2 \times R} d^3 x
\epsilon^{\mu\nu\lambda}(\frac{1}{2}V_{\mu}\partial_{\nu}V_{\lambda}
+\frac{i}{3}V_{\mu}V_{\nu}V_{\lambda}-\frac{1}{2}\hat{V}_{\mu}\partial_{\nu}\hat{V}_{\lambda}
+\frac{i}{3}\hat{V}_{\mu}\hat{V}_{\nu}\hat{V}_{\lambda})
\end{eqnarray}
Here we adopt the formalism where $SU(4)_R$ symmetry is manifest. We
have also introduced the conformal mass term for the scalars and made it equal to $1/4$. This sets the radius of the sphere to one (volume equal to $4\pi$).

The equation of motion  of the gauge fields are given by
\begin{eqnarray}
\frac{k}{2\pi} \epsilon^{\m\n\l}F_{\n\l}&=&i\bar{\phi}_AD_{\m}\phi^A-iD^{\m}\bar{\phi}_A
\phi^A  \label{eq1} \\
-\frac{k}{2\pi} \epsilon^{\m\n\l}\hat{F}_{\n\l}&=&i\phi^AD_{\m}\bar{\phi}_A-iD^{\m}\phi^A
\bar{\phi}_A.  \label{eq2}
\end{eqnarray}
We assume the gauge group to be  $U(N) \times U(N)$.
We turn on the gauge flux $s_i$ on the first gauge group
\begin{equation}
F_{ii}=\frac{s_i}{2} \sin\theta d\theta \wedge d\phi  \,\, i=1\cdots N  \label{fii}
\end{equation}
for the i-th diagonal component so that
\begin{equation}
\frac{1}{2\pi} \int _{S^2}  F_{ii}=s_i.
\end{equation}
is integer (we are ignoring the $1/k$ fractional flux, as we will set  most of the $s_i\to 0$ generally).

This is the properly normalized Dirac quantization condition. It makes specific use of the charges of the particles that are feeling the magnetic field configuration.
To satisfy the Gaussian constraints of the first gauge group we choose
\begin{equation}
\phi^1_{ii}=  a_i exp (\frac{it}{2})   \label{eq3}
\end{equation}
then solving the equations of the gauge field we find that  $a_i=\sqrt{\frac{ks_i}{4\pi}}$. We are considering only half BPS configurations. (This is similar to the analysis of \cite{JS}).

On the other hand, this gives nontrivial flux to the second gauge
group so that one has to choose
\begin{equation}
\hat{F}_{ii}=\frac{s_i}{2} \sin\theta d\theta \wedge d\phi
\end{equation}
,finding, as expected, that we need to match the same gauge fluxes for both gauge groups.
This makes the gauge group to be reduced to $U(1)^{N} \times \hat{U}{(1)}^{N}$ in genera.

If we look for the quadartic fluctuations, the scalar fields
$\phi^A_{ij}$ with $A=2,3,4$ pick up
a mass term due to the vev of the scalar $\phi^1$. The analysis of the mass matrix between two brane locations on the moduli space was done in \cite{BT}.
Here, we need to substitute the vevs of $\phi^1$ into that computation. The result for connecting eigenvalue $i$ and eigenvalue $j$ is proportional to
\begin{equation}
m^2\sim ||\phi^1_{ii}|^2-|\phi^1_{jj}|^2|^2\sim (s_i-s_j)^2
\end{equation}
If we include all of the  normalization factors we find that the lagrangian for the transverse $\phi_A$  quadratic fluctuations is
\begin{equation}
Tr(-D_{\mu} \bar{\phi}_A D^{\mu} \phi^A -\frac{1}{4} \bar{\phi}_A \phi^A)
-(\frac{s_i-s_j}{2})^2\sum_{A=2,3,4}
(|\phi^A_{ij}|^2+|\phi^A_{ji}|^2))+\cdots
\end{equation}
where
\begin{equation}
D_{\mu}\phi^A_{ij}=\partial_{\mu}
\phi^A_{ij}-i(\bar{V}_{\mu \,\,ii}-\bar{V}_{\mu \,\, jj})\phi^A_{\mu \,\,ij}.  \,\,\, A=2,3,4
\end{equation}
Here $d \bar{A}_{ii}= F_{ii}$ of eq. (\ref{fii}).

Thus $\phi^A_{ij}$ feels the monopole background with the charge $\frac{s_i-s_j}{2}$.
Upon the dimensional reduction on $S^2$ the relative ratio of the
kinetic term and the mass terms are not changed so that we have the
contribution
to the mass term from the scalar potential
$m_1^2=(\frac{s_i-s_j}{2})^2$.
One can trace  the origin of this mass term from the previous $N=2$
formalism.
This mass term comes from the commutator term with the auxiliary
field $\sigma$ in the scalar potential (\ref{scalarpotential}) since
$\sigma$ has the nontrivial value due to the vev of $\phi^1_{ii}$.
On the other hand on the monopole background with the flux $s_i-s_j$
we have the contribution
$m_2^2=l(l+1)-(\frac{s_i-s_j}{2})^2+\frac{1}{4}$
where the first two terms come from the angular momemtum operator
while the last term comes from the conformal mass term. These are the monopole spherical harmonic energies for a scalar field in the presence of a magnetic field (see \cite{WuYang}).

Thus the total energy of the fluctuation is given by
\begin{equation}
E=\sqrt{m_1^2+m_2^2}=l+\frac{1}{2}.  \label{ms}
\end{equation}
The allowed value of $l=\frac{|s_i-s_j|}{2}, \frac{|s_i-s_j|}{2}+1, \cdots$ and $l$ is
the total angular momentum of the excitation. Note that the dependence
of the energy
on the monopole background with the charge $\frac{s_i-s_j}{2}$
cancels with the mass term from the scalar vevs.\footnote{Such
cancellation is also  observed at \cite{Kim:2009wb}.}
Also, these fluctuations can have $R$-charge equal to $\pm 1/2$, which depends on the decomposition of the fields $\phi^{A}$ into R-chare multiplets. Both values arise: the $B^\dagger$ have R-charge $-1/2$, while the $A$ have R-charge $1/2$. Similarly for
their complex conjugate quanta.
If one wants to saturate
the BPS bound for a particular ${\cal N}=1$, one is only allowed to use quanta of $A,B$.

Note that  $\phi^A_{ii} \,\,A=2,3,4$ component does not have any monopole
background dependence in the covariant derivative as well as in the
mass term due to the scalar potential. Hence we have the same energy
expression
\begin{equation}
E=l+\frac{1}{2}
\end{equation}
but allowed values of $l=0,1,\cdots$ are different. One can show that
this holds for $\phi^1_{ii}$ as well. This justifies the
eq. (\ref{eqA}) when we discuss the dual configurations of the fuzzy
sphere giant.
It's trickier to figure out the spectrum of $\phi^1_{ij}$ for $i\neq
j$ since there's nontrivial mixing between the gauge field and scalar
component. With the absence of the monopole  background such analysis
is carried out at \cite{BT}.  But the above
results are sufficient for  the later discussions \footnote{This mixing analysis has been done in \cite{KimMadhu}, after our original results were released}.

It is straightforward to work out the energy of the fermions in the
monopole background using the results derived in \cite{Kapustin}.
For the diagonal component $\Psi^A_{ii}$ with $A=1,2,3,4$
\begin{equation}
E=j+\frac{1}{2}
\end{equation}
where allowed $j=\frac{1}{2}, \frac{3}{2} \cdots $. These components do
not see the presence of the monopole background, similar to the diagonal
components of scalar fields.
For the off-diagonal modes $\Psi^A_{ij}$ with $A=1,2,3,4$ we have to solve the Dirac
equation in the presence of monopole background with charge
$\frac{s_i-s_j}{2}$.
Again the angular momentum operator gives rise to the contribution
\begin{equation}
E_1=\sqrt{(j+\frac{1}{2})^2-(\frac{s_i-s_j}{2})^2}
\end{equation}
while the Yukawa terms give the additional contribution
\begin{equation}
E_2=|\frac{s_i-s_j}{2}|
\end{equation}
This is related by supersymmetry to the mass of the bosons we found above.
Hence the total energy contribution is
\begin{equation}
E_t=j+\frac{1}{2}
\end{equation}
The allowed values of $j$ is $j=|\frac{s_i-s_j}{2}|-\frac{1}{2},
|\frac{s_i-s_j}{2}|+\frac{1}{2} \cdots$.

Since we get a specific mass term in the lagrangian, the fermions are
necessarily paired (they are not massless)
and there are no zero modes.\footnote{Note that
$j=|\frac{s_i-s_j}{2}|-\frac{1}{2}$ is not allowed if $s_i-s_j=0$,
thus avoiding the zero modes.} This means that there are no difficulties in defining the charges of the ground states and the classical result for the charges of the semiclassical states is exact: there is a standard Fock space of states.

Now utilizing the above results, one can construct the dual states for a degeneration limit of
the M2 giant torus state. Suppose we just turn on the magnetic flux
$M$ of $F_{11}$ so that
\begin{equation}
kM=\phi_{11}^1.   \label{Flux}
\end{equation}
We assume that $M >> 1$ and can treat the resulting configuration
classically.
The energy due to the presence of $\phi^1_{11}$ quanta is given by
\begin{equation}
E=\frac{kM}{2},   \,\, M > 0.  \label{E1}
\end{equation}
Now consider a charged excitation in the presence of the monopole
background
of the above setup. Say, we have one excitation of $\phi_{12}^2\sim A^2_{12}$ which has
the energy $E=\frac{M+1}{2}$ and another excitation of $\phi_{21}^3\sim B^1_{21}$
which has the same energy and R-charge as  $A_{12}^2$. On $S^2$ we cannot have
single charged excitation and we need a pair of charged excitation
with opposite charges. Thus the minimum energy one can have is
\begin{equation}
E=M+1.    \label{E2}
\end{equation}
This is related to the gauge invariance of the resulting
operators. However, these also carry angular momentum.

To saturate the BPS bound, we need to maximize the angular momentum between the two quanta.
For states in a large magnetic field,  quanta localize into landau level states. The angular localization is such that each such state occupies $1/M$ of the area of the sphere. The angular momentum is directed along the position of the charged particle, with a sign that indicates its charge.

A field $A_{12}$ with $L_z=(s_1-s_2)/2$ ends up in the north pole, with a field $B_{21}$ with the same $L_z$ ends up in the south pole: they have opposite charges with respect to the background magnetic field.

The total energy of the resulting configuration is the sum of
(\ref{E1}) and (\ref{E2}) so that
\begin{equation}
E_t=\frac{kM}{2}+M+1 \sim
\frac{kM}{2}+M=kM(\frac{1}{2}+\frac{1}{k})=J(\frac{1}{2}+w)  \label{Et}
\end{equation}
where $J=kM$ and $w=\frac{1}{k}$. Note that string charge is
quantized in units of $\frac{1}{k}$ because of the orbifolding
procedure. Thus eq. (\ref{Et}) precisely matches with the energy
relation of the M2 giant torus eq. (\ref{etorus}). One might
recall that in the presence of the monopole background with charge
$\frac{M}{2}$ one  particle with minimal charge gives rise to the
angular momentum $\frac{M}{2}$ while one particle with opposite
charge in the opposite polar position has the same contribution so
that a pair of charge particles give rise to the angular momentum
$M$. Thus the above configuration dual to M2 giant torus nicely
fits with the BPS formula $E=R+J$ with $R=\frac{kM+2}{2}$ and
$J=M$.

One might note that in the above construction we choose scalars of
three different $U(1)$ charges out of $SU(4)_R$ charges, the
diagonal $\phi^1$ and off-diagonal $\phi^2, \phi^3$. However one
can have different field theory configuration with the same BPS
formula in the free theory limit. In \cite{KimMadhu} the
fluctuation spectrum of $\phi^1_{1i}, \phi^1_{i1}$ is analyzed in
the presence of the magnetic flux and the vev of $\phi^1_{11}$  in
eq. (\ref{Flux}). If we choose the gauge where there are no gauge
field excitations,   there are fluctuating modes of $\phi^1_{12}$
having energy $\frac{M+1}{2}$.\footnote{ In \cite{KimMadhu}, they
work out gauge field excitations which are gauge equivalent to the
scalar field excitations of $\phi^1_{12}$. However the Hamiltonian
expression eq. (\ref{energy}), used in the evaluation of the
energy of the scalar field excitations, as well as R-charge
expression eq. (\ref{Rcharge}) is gauge invariant. }Thus one can
have the same energy configuration of eq. (\ref{Et}) using
$\phi^1_{11}$ of eq. (\ref{E1}) and $\phi^1_{12}$, $\phi^3_{21}$
of energy $\frac{M+1}{2}$.\footnote{However, $\phi^1_{12},
\phi^1_{21}$ are lacking the modes with energy $-\frac{M+1}{2}$
and $\frac{M+1}{2}$ respectively. On the other hand, $\phi^1_{12}$
has two modes with energy $\frac{M+1}{2}$.} In this configuration
we choose scalars of two different $U(1)$ charges. It is noted in
\cite{NT} that in the gravity side the configuration with two
$U(1)$ charges is BPS while the configuration with three $U(1)$
charges is not. This suggests that as we turn on the interaction
the configuration with two $U(1)$ charges remain BPS while three
$U(1)$ configuration gets non-BPS. Note that our derivation of the
BPS formula is made at the free theory limit and to compare with
the gravity side we have to work out the effect of the
interactions.

Notice that when we did the computation we found an extra mass term contribution from the scalar vevs that  was positive. Without it, the corresponding quark quanta would have energies below the unitarity bound: this is  especially noticeable for large fluxes. The reason is that the ground state of a charged particle in a magnetic field has large angular momentum. While the magnetic monopole spherical harmonics barely account for the lowest Landau level localization energy frequency $w^2 \sim P^2\sim [\hbar/\delta X]^2\sim M$ due to the uncertainty principle. The extra mass  term restores the unitarity relation
\begin{equation}
\delta E\geq \delta R+ \delta L_z
\end{equation}
for fluctuations. The background already satisfies
\begin{equation}
E_{back}= R_{back}
\end{equation}

This shows us that we can get one string on a fuzzy sphere dual that stretches between the north and south pole and such that it saturates the BPS bound. Finding one string, we can add many. In the weak coupling limit these don't interact. When the number of strings grows sufficiently, one can not ignore backreaction any longer. It would be interesting to see if one can show how the full giant torus is recovered. Since these states saturate the unitarity bound, the quantum corrections must be positive or zero. It would be interesting to check if these configurations stay BPS after we include interactions.

\section{Other more general BPS states}

We have completed the basic analysis of configurations that can in principle give rise to the giant torus. Notice that the quanta we added need not be in the lowest Landau level. One can still consider a quark and an
antiquark in different Landau levels. Again, we can saturate the BPS bound if we maximize $L_z$ between them. This gives rise to a large degeneracy of possible BPS states. Since the quanta can exchange angular momentum via the unbroken $U(N-1)\times U(N-1)$ interactions, we expect that this degeneracy gets mostly lifted by quantum corrections of 'spin chain' type. Perhaps one such state survives for each angular momentum.
Classically, going to higher landau levels corresponds to wider synchroton orbits of a particle in a magnetic field. Thus the quark and antiquark get displaced from the north/south poles and eventually can be described by a classical circular trajectory around the pole. The possible exchange of momentum suggests that the minimum energy configuration with fixed angular momentum corresponds to the quark and antiquark being antipodal to each other, so each of them ends up with the same average angular momentum. Notice that a big factor in the stabilization of these configurations is played by angular momentum. The R-charge can not disappear either.

Notice also that our analysis applies not just in the case of one monopole flux, but that we can make general excitations about half-BPS configurations with many such fluxes. These configurations
have `strings' stretching between many pairs of fuzzy spheres. They can begin at the north pole of the first and end at the north pole of the second (so that they have classically minimal mass), and this produces the correct angular momentum by subtracting the angular momentum between the ends, as appropriate for the charge of the state. With the angular momentum localization on the sphere, we find that these are also BPS fluctuations. To compensate for the charge of the states, there must be antiquarks stretching between the south poles of these objects, but these pairings can be different in general.
There are in principle a lot of such states that stretch between various D-branes. This allows us to build a large degeneracy of BPS states (at least classically), and suggests a mechanism for counting general $1/16$ BPS black holes. In the case of $AdS_5$, black holes require spin \cite{GReall}. Generally, $1/8$ BPS states as those of the chiral ring are too few to produce classical black hole states. In the case of $\CN=4$ SYM theory this was shown in \cite{BlargeN}, because all BPS states are diagonal. Thus their entropy grows as $N$ and not $N^2$. Attempts to count the more general degeneracies of black holes from CFT have been made for $AdS_5$  in \cite{BRS}, see also \cite{Grant:2008sk}, but the results are either in free theory, or the configurations that are understood are all abelian. A complete set of references on this literature and a discussion of the issues can be found in \cite{BBSM}.

In our case we seem to be able to do better. We have large collections of D-branes and the strings stretched between them are BPS. This gives extra factors of $N$ easily.
This suggests that we have enough states to produce a big enough entropy to account for the black hole micro-state counting. Such a counting is beyond the scope of the present paper as it would require checking that the states saturate the BPS bound at the next order in perturbation theory.

Notice also that if we follow the same logic for a situation where  two monopoles are not half BPS, then the mass term contribution from the scalar backgrounds becomes proportional to
\begin{equation}
m_{ij}^2 \sim |s_i+s_j|^2 -4 s_is_j \cos\vartheta^2> |s_i-s_j|^2
\end{equation}
This follows from eq. (66) in \cite{BT}. Thus, if the monopoles are not aligned in R-charge, the `stretched string' states are not BPS any longer. This may suggest that the BPS nature of these stretched string states is rather fragile. Any argument about their BPS properties once interactions are included needs to be done very carefully.

\section*{Acknowledgements} D. B. would like to thank Igor Klebanov , Juan Maldacena and Mauricio Romo for discussions related to this work.
D.B.  is supported in part by DOE under grant DE-FG02-91ER40618.
J.P. is  supported in part by the KOSEF SRC Program through CQUeST
at Sogang University, by KOSEF Grant R01-2008-000-20370-0,  by the
National Research Foundation of Korea(NRF) grant funded by the
Korea government(MEST) with the grant number 209-0085995
 and
by the Stanford Institute for Theoretical Physics.

\end{document}